\documentclass[twocolumn,showpacs,amsmath,amssymb,prb]{revtex4}

\usepackage{graphicx}
\usepackage{dcolumn}
\usepackage{bm}

\def\be{\begin{eqnarray}}
\def\ee{\end{eqnarray}}
\def\ba{\begin{array}}
\def\ea{\end{array}}

\begin{document}

\preprint{1}

\title{Incompressible strips in dissipative Hall bars as origin of 
quantized Hall plateaus}

\author{Afif Siddiki}%
\author{Rolf R. Gerhardts}
\affiliation{Max-Planck-Institut f\"ur Festk\"orperforschung, 
Heisenbergstrasse 1, D-70569 Stuttgart, Germany}%

\date{\today}

\begin{abstract}
We study the current and charge distribution in a two dimensional electron 
system, under the conditions of the integer quantized Hall
effect, on the basis of a quasi-local transport model, that includes
non-linear screening effects on the conductivity via the
self-consistently calculated density profile. The existence of
``incompressible strips'' with integer Landau level filling factor is
investigated within a Hartree-type approximation, and non-local
effects on the conductivity along those strips are simulated by a
suitable averaging procedure. This allows us to calculate the Hall and the
longitudinal resistance as continuous functions of the magnetic field $B$,
with plateaus of finite widths and the well-known, exactly quantized
values. We emphasize the close relation between these plateaus and the
existence of incompressible strips, and we show that for $B$ values
within these plateaus the potential variation across the Hall bar is
very different from that for $B$ values between adjacent plateaus, in
agreement with recent experiments.
\end{abstract}

\maketitle

\section{\label{sec:1} Introduction}

The application of the quantized Hall effect (QHE) \cite{vKlitzing80:494} 
as resistance standard, and its importance for modern metrology,
relies
on the extremely high reproducibility (better than $10^{-8}$) of
certain quantized resistance values. \cite{Bachmair03:14}
This extreme reproducibility points to an universal origin, which is
independent on special material or sample properties. 
The purpose of the present paper is to propose and evaluate a
quasi-local transport model that allows us to calculate, first, the
potential and current distribution in a two-dimensional electron system (2D ES)
under the conditions of the QHE, and, second, the longitudinal
and the Hall resistance, $R_l(B)$ and  $R_H(B)$, in the plateau
regimes of the QHE and in between. Whereas the resistance values
between the QH plateaus will depend on details of the used
conductivity model, the exactly quantized plateau values result from
the existence of sufficiently wide ``incompressible strips'' along
which the local conductivity vanishes, since occupied and unoccupied
states are separated by an energy gap (Landau quantization).
Localization assumptions,\cite{Kramer03:172} which played an important
role in early approaches to the QHE, are not included in our model.
Our model is motivated by  a recent experimental investigation of the
Hall-potential in a narrow Hall bar, \cite{Ahlswede01:562}
and a critical reexamination of a subsequent model
calculation. \cite{Guven03:115327}

Experimental information about the actual current and potential
distribution in a Hall bar under QHE conditions has been obtained  recently
by Ahlswede and coworkers
\cite{Weitz00:247,Ahlswede01:562,Ahlswede02:165}  with a scanning
force microscope \cite{Weitz00:349}.  
The data were interpreted in terms of ``incompressible strips'' with
constant electron density (corresponding to the filling of an
integer number of Landau levels),
\cite{Chklovskii92:4026,Chklovskii93:12605,Lier94:7757,Oh97:13519}  
which are expected to develop in an inhomogeneous  2D ES 
as a consequence of its strongly non-linear low-temperature
screening properties \cite{Siddiki03:125315} in a strong perpendicular
magnetic field. 
If the filling factor in the center of the sample was
slightly larger than an integer, the Hall potential was found to drop
completely across two strips, while being constant elsewhere. With
decreasing $B$, the strips moved towards the sample edges, just as one
expects for the incompressible strips in a sample, in which the
electron density decreases gradually from a maximum in the center
towards the edges.
 If the center filling factor was slightly below an integer,  a
gradual potential variation  was observed, either linear or a non-linear,
 without clear indication for incompressible strips.\cite{Ahlswede01:562} 

This interpretation was supported by
subsequent theoretical work of G\"uven and Gerhardts (GG)
\cite{Guven03:115327}, who extended the self-consistent
Thomas-Fermi-Poisson approximation (TFPA)
\cite{Lier94:7757,Oh97:13519,Siddiki03:125315}
for the calculation of  electron density profile and electrostatic
potential to a non-equilibrium situation with  a
position-dependent electrochemical potential, determined by the
presence of an applied dissipative current through the sample. 
Electrochemical potential and current density was calculated  from  a local
version of Ohm's law, with a local model for the conductivity tensor, 
determined by the position-dependent electron density. The feed-back
of the current distribution on electron density and the measurable
potential profile  \cite{Ahlswede01:562} was included by the
assumption of local equilibrium in the stationary non-equilibrium situation.

In agreement with the experiment \cite{Ahlswede01:562}, the
calculation  \cite{Guven03:115327} shows a linear variation of the
Hall potential across the sample if there are no incompressible
strips, e.g. for sufficiently high temperature or if the magnetic
field is so strong, that the local filling factor is everywhere in the
sample less than two (spin-degeneracy is assumed and interactions
 which might lead to the fractional quantized Hall effect are neglected). 
Also for center filling factors slightly larger than 2 or 4 the
calculation confirms the experiment,  showing that the potential drops
across broad incompressible strips and is constant elsewhere.
However, due to the use of the TFPA, GG\cite{Guven03:115327} obtain
incompressible strips whenever the center filling factor is larger
than 2,\cite{Chklovskii92:4026,Chklovskii93:12605,Lier94:7757,Oh97:13519}    
and due to the strictly local conductivity model these dominate the
current and potential distribution, and lead to  vanishing
longitudinal resistance. Thus,  the model assumptions of
Ref.\onlinecite{Guven03:115327} lead to serious disagreement with
important aspects of the experiment. 

The purpose of the present paper is to improve on the model
of GG\cite{Guven03:115327} so that, first, 
qualitative agreement between the calculated and the measured
potential distribution is achieved  for all filling factor regimes,
and, second, reasonable results for $R_l(B)$ and $R_H(B)$ are obtained.
 Following the
lines suggested by GG\cite{Guven03:115327}, we investigate in
Sec.~\ref{sec:Hartree} the conditions for the existence of
incompressible strips, using a Hartree approximation. In
Sec.~\ref{sec:Conduc} we reexamine 
and weaken the strictly local conductivity model, and show that a
simple spatial-averaging procedure of the local conductivities can
simulate  corrections expected from a Hartree
calculation for the equilibrium state and from a non-local transport
calculation. Transport results based on the self-consistent Born
approximation will be presented and discussed in
Sec.~\ref{sec:Discuss}.
In the present work we will restrict our consideration on the linear
response regime, so 
that heating effects, which might destroy incompressible strips in the
presence of high currents,  \cite{Guven03:115327} can be neglected.

\section{\label{sec:Hartree} 
Existence of incompressible strips} 
\subsection{\label{sec2:1} Electrostatic self-consistency}
Following  Ref.\onlinecite{Guven03:115327}, we consider a 2D
ES in the plane $z=0$, with translation invariance in the $y$
direction and an electron density $n_{\rm el}(x)$ confined to the
interval $-d < x <d$. The confinement potential $V_{bg}(x)$ is
determined by fixed background charges and boundary
conditions on metallic gates. The mutual Coulomb interactions between
the electrons are treated in a Hartree-type approximation, i.e., are
replaced by a potential energy term $V_H(x)$ which is
determined via Poisson's equation by the electron density. Exchange
and correlation effects 
are neglected, and spin degeneracy is assumed. Thus, the electrons move
in an effective potential 
\be
&& V(x)= V_{bg}(x) + V_H(x), \label{eq:Vvon_x}\\
&& V_H(x)= \frac{2e^2}{\bar{\kappa}} \int_{-d}^d dx'\, K(x,x') n_{\rm el}(x'),
\label{eq:V_Hartree}
\ee
where $-e$ is the charge of an electron, $\bar{\kappa}$ an average
background dielectric constant, and the kernel $K(x,x')$ solves
Poisson's equation under the given boundary conditions. 
Kernel and background potential for the frequently used model,
\cite{Chklovskii92:4026,Chklovskii93:12605,Lier94:7757,Oh97:13519}
that assumes all charges and gates to be confined to the plane $z=0$,
are taken from Ref.\onlinecite{Guven03:115327},
 \be K(x,x')=\ln
\left|\frac{\sqrt{(d^2-x^2)(d^2-x'^2)}+d^2-x'x}{(x-x')d} \right| \,,
  \label{eq:kernel-inplane} \\
 V_{bg}(x)=-E_{bg}^0\sqrt{1-(x/d)^2} \,, \quad E_{bg}^0=2 \pi
e^2n_{0}d/\bar{\kappa} \,,
\label{eq:backgroundpot}
 \ee
where $en_0$ is the homogeneous density of positive background charges
in the Hall bar.  Other meaningful and
tractable boundary conditions are also possible.\cite{Siddiki03:125315}

To perform explicit calculations,  one needs a prescription
to calculate the electron density for given effective potential
$V(x)$, which then  together with Eqs.~(\ref{eq:Vvon_x}) and
(\ref{eq:V_Hartree}) 
 completes the electrostatic self-consistency. The self-consistent TFPA
\cite{Lier94:7757,Oh97:13519,Guven03:115327,Siddiki03:125315} 
takes this prescription from the
Thomas-Fermi approximation (TFA)
\be \label{eq:TFA}
n_{\rm el}(x)=\int dE \, D(E)\, f(E+V(x)-\mu^{\star}),
\ee
with $D(E)$ the density of states (DOS), $f(E)=1/[\exp(E/k_BT)+1]$ the
Fermi function, $\mu^{\star}$ the electrochemical potential (being
constant in the equilibrium state), $k_B$ the Boltzmann constant, and
$T$ the temperature.

\subsection{Hartree approximation}
A less restrictive approximation would be to insert $V(x)$ into
Schr\"odinger's equation,
\be \label{eq:Schr2D}
\left[\frac{1}{2m}\left({\bf p}+\frac{e}{c}{\bf
    A}\right)+V(x)\right]\Phi_{\lambda}({\bf r})=
E_{\lambda}\Phi_{\lambda}({\bf r}) ,
\ee
with ${\bf A}({\bf r})$ a vector potential describing the magnetic
field ${\bf B}=(0,0,B)=\nabla \times {\bf A}$, and to calculate the
density from the eigen-energies $E_{\lambda}$ and -functions
$\Phi_{\lambda}({\bf r})$,
\be \label{eq:density_gen}
n_{\rm el}({\bf r})=\sum_{\lambda} | \Phi_{\lambda}({\bf r})|^2
f(E_{\lambda}- \mu^{\star}).
\ee
Exploiting the symmetry of our system, we may use the Landau gauge,
${\bf A}({\bf r})=(0,Bx,0)$, and factorize the wavefunctions,
$\Phi_{\lambda}({\bf r})=L_y^{-1/2}\exp(iky)\phi_{n,X}(x)$, with $L_y$
a normalization length,  $X=-l^2k$
a center coordinate, $l=\sqrt{\hbar/m\omega_{c}}$ the magnetic length,
and $\omega_{c}=eB/(mc)$ the cyclotron frequency. The Schr\"odinger equation
then reduces to
\be 
\left[ -\frac{\hbar^{2}}{2m} 
\frac{d^{2}}{dx^{2}}+\frac{m}{2}\,\omega_{c}^{2}(x-X)^{2}
+V(x)  \right]\phi_{n,X}(x)=\nonumber\\
 E_n(X) \phi_{n,X}(x),
\label{eq:ham}
\ee
and the electron density becomes 
\be n_{\rm el}(x)=\frac{g_{s}}{2\pi l^2} 
\sum_{n}\int dXf(E_{n}(X)-\mu^{\star})|\phi_{n,X}(x)|^{2}, \label{eq:geneled}
\ee
where $g_{s}=2$ takes the spin degeneracy into account and the sum
over $X$ has been replaced by an integral, $L_y^{-1}\sum_X \Rightarrow (2\pi
l^2)^{-1} \int dX$.


\subsection{Thomas-Fermi Approximation (TFA) \label{sec:2-tfa}}
If the potential $V(x)$ varies slowly on the scale of the magnetic
length $l$, its effect on the lowest Landau levels  (LLs) can be treated
perturbatively, with the lowest order result
\be \label{eq:TFA-energies}
E_n(X) \approx E_n +V(X), \quad E_n=\hbar \omega_c(n+1/2).
\ee
On the length scale relevant for the variation of $V(x)$,
the extent of the  Landau wavefunctions may be neglected, 
 $|\phi_{n,X}(x)|^{2} \approx \delta(x-X)$. Then
 the Hartree result for the electron density,
Eq.~(\ref{eq:geneled}), reduces to the TFA,
Eq.~(\ref{eq:TFA}), with the Landau DOS
\be \label{landau-dos}
D(E)=\frac{1}{\pi
l^{2}}\sum_{n=0}^{\infty}{\delta (E-E_n)}.
 \ee

To evaluate the self-consistent  TFPA we follow 
Ref.\onlinecite{Guven03:115327}. First we fix the sample width $2d$
and the density of 
positive background charges $n_0$, and thereby according to
Eq.~(\ref{eq:backgroundpot}) the background potential $V_{bg}(x)$ and
the relevant screening parameter  $\alpha_{sc}\equiv \pi a_0/d $, with 
$a_0=\bar{\kappa}\hbar^2/(2m e^2)$ the screening length ($2 a_0=
9.79\,$nm for GaAs). Next we choose the actual width
$2b$ of the density profile at $T=0$ and $B=0$, and solve for $|x|\leq
b$ the linear integral equation 
\cite{Guven03:115327} 
\be \label{eq:linint}
V(x)\!-\!V_{bg}(x)\!=\!\frac{1}{\alpha_{sc}}\int_{-b}^{b}\! \frac{dx'}{d} \,
K(x,x')\,[\mu_0^{\star}\!-\!V(x')],
\ee
[with $\mu_0^{\star}\!=\!V(\!-\!b)\!=\!V(b)$]
to which the self-consistent TFPA reduces in this limit.
From the corresponding density profile $n_{\rm el}(x;B=0,T=0)=D_0
[\mu_0^{\star}-V(x)]$, with  $D_0=m/(\pi \hbar^2)$ the DOS of the
2D ES at $B=0$, we calculate $V(x)$ for $|x|\leq d$, the average density
$\bar{n}_{\rm el}=\int_{-d}^d dx\, n_{\rm el}(x;B\!=\!0,T\!=\!0)/2d$ with the
corresponding Fermi energy $E_F=\bar{n}_{\rm el}/D_0$, and, for later
reference, the Fermi energy $E_F^0= n_{\rm el}(0;B\!=\!0,T\!=\!0)/D_0$
corresponding to the electron density at the center.

In the following we will consider only symmetric density profiles and
take $b$, or equivalently the depletion length $d-b$, as a 
free parameter, that fixes the density profile and the
electro-chemical potential $\mu_0^{\star}$ at $B=0$ (where the
temperature dependence is weak). In real samples $\mu_0^{\star}$ may be
determined by an electron exchange between the 2D ES and its
surrounding, which may be possible at high but not at low
temperatures. A restriction that fixes $\mu_0^{\star}$ will also determine
the value of $b$.

Next, we fix the value of the magnetic field and start with a high
temperature to calculate the electron density from Eqs.~(\ref{eq:TFA})
and (\ref{landau-dos}) self-consistently with
Eq.~(\ref{eq:V_Hartree}), using the previously calculated potential
$V(x)$ as initial value. Finally we lower $T$ stepwise
until the required low temperature is reached, and iterate (using a
Newton-Raphson approach) at each temperature until convergence is achieved.

\begin{figure}[h]
{\centering
  \includegraphics[width=0.8\linewidth,angle=-90]{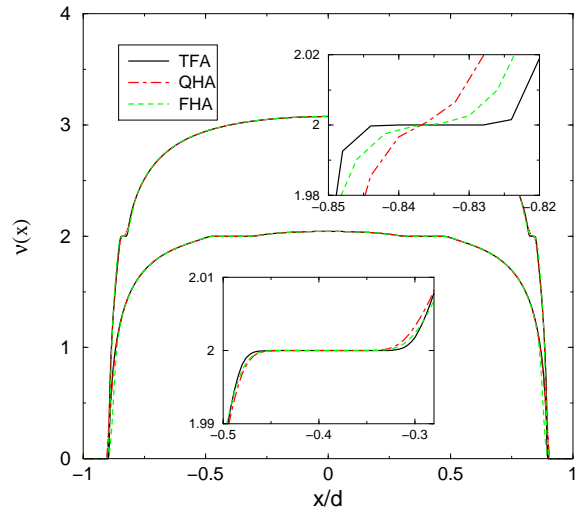}
\caption{ \label{fig:incomp_compare}
Electron density profiles for two values of the magnetic field ($\hbar
\omega_c/E_F^0= 0.94$ and $0.65$) and
different approximations: Thomas-Fermi (solid lines), Hartree
(dashed), and quasi-Hartree (dash-dotted). The insets show the enlarged plateau
regions for both cases. $\alpha=0.01$,  $k_BT/E_F^0=0.0124$. Kinks in
the upper inset indicate mesh points.
 }}
\end{figure}
The solid lines of Fig.~\ref{fig:incomp_compare} show results  for
$d=1.5\,\mu$m and $n_0=4\cdot 10^{11}\,$cm$^{-2}$ 
(which implies  $E_{bg}^0=4.38\,$eV) obtained for 501 equidistant mesh
points, $-d=x_0<x_n <x_N=d $ ($N=500$).  The density profile was fixed
by choosing $b/d=0.9$, which yields $\bar{n}_{\rm el}=2.9
\cdot 10^{11}\,$cm$^{-2}$ and $n_{\rm el}(0;B\!=\!0,T\!=\!0)=3.58
\cdot 10^{11}\,$cm$^{-2}$, and thus, 
with  $D_0=2.8\cdot 10^{10}\,$meV$^{-1}$cm$^{-2}$ for AlGa,
 $E_F= 10.37\,$meV and $E_F^0= 12.75\,$meV. We prefer to use  $E_F^0$
(rather than $E_{bg}^0$) as an reference energy, since it has the same
order of magnitude as the cyclotron energies of interest.
For finite $B$ we define an effective center filling factor
$\nu_0=2\pi l^2 n_{\rm el}(x\!=\!0;B\!=\!0,T\!=\!0)=2 E_F^0/\hbar
\omega_c$.

 The result obtained for the TFPA (solid lines in
 Fig.~\ref{fig:incomp_compare}) shows the expected well  
developed incompressible strips with constant electron density at local
filling factor $\nu(x)=2$. For the larger $B$ value we obtain wide
density plateaus at $0.32\leq |x|/d \leq 0.46$, in each of which we
find at 36 $x_n$ values, with high accuracy, $\nu(x_n)=2$. For the lower
$B$ value the incompressible strips are much narrower, however we
obtain the high precision values $\nu(x_n)=2$ still at five
neighboring $x_n$ values.
\begin{figure}[h]
{\centering
  \includegraphics[width=.95\linewidth]{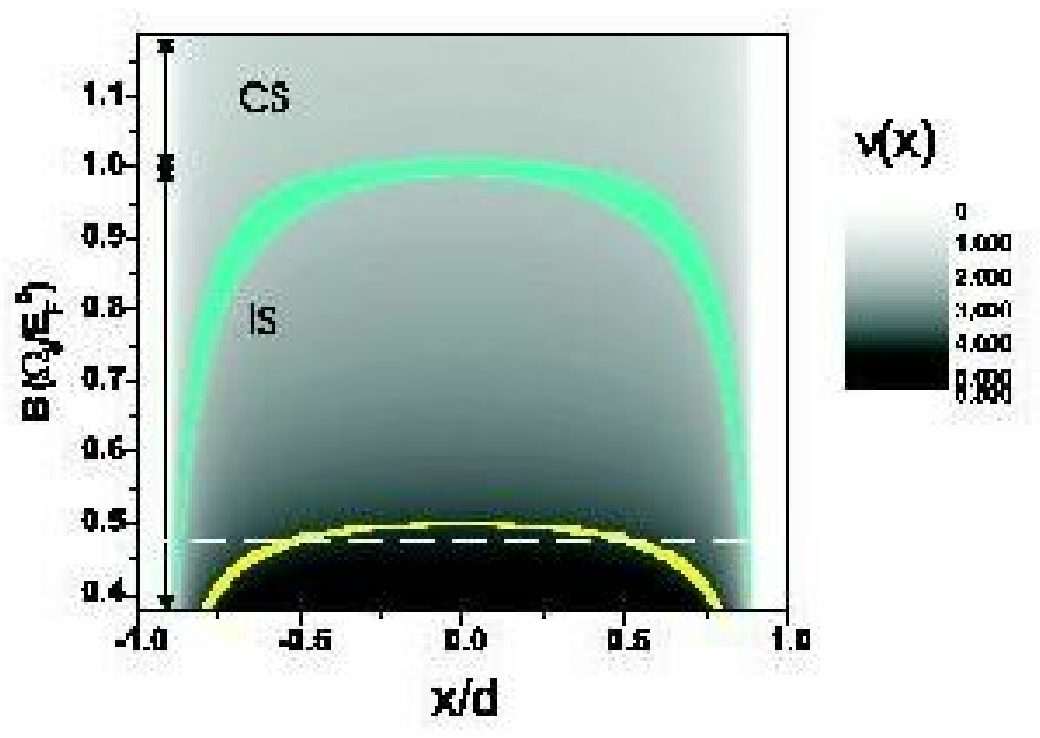}
\caption{ \label{fig:sum-TFA} (color online)
Gray scale plot of filling factor versus position $x$ and magnetic
field $B$, calculated within the TFPA. The regions of incompressible strips
with $\nu(x)=2$ and $\nu(x)=4$ are indicated. For sufficiently large
$B$ ($\Omega_c\equiv \hbar \omega_c > E_F^0$) the system is
compressible (indicated by ``CS''), while for the lower $B$ values
included in the figure it always contains incompressible strips (``IS'').
The dashed horizontal line refers to Fig.~\ref{fig:nu-sh-sl}
below; $\alpha=0.01$, $k_BT/E_F^0=0.0124$. 
 }}
\end{figure}
Typical results of the TFPA are summarized in Fig.~\ref{fig:sum-TFA},
which shows, as a gray scale plot, the filling factor profile for
varying magnetic field, with horizontal lines corresponding to a fixed
$B$ value. At sufficiently large $B$ field, the local
filling factor $\nu(x)$ is everywhere in the Hall bar less than 2,
and the 2D ES is completely compressible. At somewhat lower $B$
($\hbar \omega_c/E_F^0\approx 1$) the center of the sample becomes
incompressible with local filling factor $\nu(x)=2$, while $\nu(x)$
gradually decreases outside the incompressible center and falls off to
zero in the depletion regions at the sample edges. With further
decreasing $B$, the filling factor in the center increases and 
incompressible strips with $\nu(x)=2$ move towards the sample edges
and become narrower.   At sufficiently low $B$ values, incompressible
strips with local filling factor $\nu(x)=4$ occur, first in the center,
and then move towards the edges. They then coexist with incompressible
strips of  $\nu(x)=2$, which exist near the edges and are narrow, but,
within the TFPA, still have a finite width. For low enough
temperature, this type of 
behavior continues at still lower values of $B$, where further
incompressible strips with successively higher filling factors evolve
from the center and move towards the edges, coexisting with the
edge-near incompressible strips of lower local filling factors.

\subsection{``Quasi-Hartree'' Approximation \label{sec:QHA}}
The dashed lines in Fig.~\ref{fig:incomp_compare} are calculated in
the Hartree approximation. We started again at $B=0$, $T=0$ and inserted
in each of the following iteration steps the previously calculated
potential $V(x)$ into the eigenvalue problem of Eq.~(\ref{eq:ham}),
took each mesh point $x_n$ as center coordinate $X$ and diagonalized
the problem in the space spanned by the eight lowest-energy
unperturbed Landau-Hermite functions
\be \label{eq:Hermite-fu}
\phi_{n,X}^0(x)=\frac{\exp(-[x-X]^2/2l^2)}{\sqrt{2^{n}n!\sqrt{\pi}l}}
\,H_{n}([x-X]/l)\,, 
\ee
where $H_{n}(\xi)$ is the $n$-th order Hermite polynomial. 
The resulting energy eigenvalues and -functions were used to
calculate the electron density according to Eq.~(\ref{eq:geneled}).
 The overall appearance of the Hartree
results for  the density
profiles and also the wide plateaus for the higher $B$ value (see
lower inset of Fig.~\ref{fig:incomp_compare}) are in
good agreement with those of the TFA. The narrow plateaus, obtained in
the TFA for the lower $B$ value, are now however smeared out. As is
clearly seen in the upper inset of Fig.~\ref{fig:incomp_compare}, the
Hartree result for the filling factor $\nu(x)$ crosses the value 2
with a finite slope.

\begin{figure}[h]
{\centering
\includegraphics[width=0.7\linewidth,angle=-90]{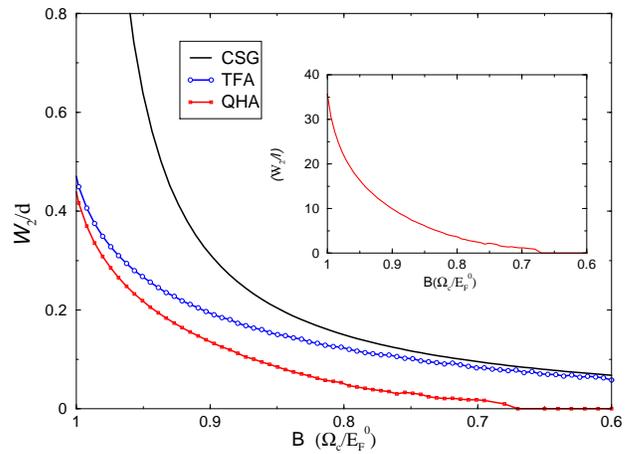}
\caption{Magnetic-field dependence of the  width of the 
$\nu=2$ incompressible strips, for three different approximations: the
  analytical result of Ref.\onlinecite{Chklovskii92:4026} (CSG), the
  TFPA (TFA), and the quasi-Hartree approximation (QHA). Note the
  inverted $B$ scale. Inset: ratio of the incompressible 
strip width to the magnetic length in QHA.  \label{comp-width-inc}}}
\end{figure}
The essential qualitative difference between the Hartree approximation
and the TFA is the neglect of the extent of the wavefunctions in the
latter. So we interpret the smearing-out of narrow incompressible
strips in the Hartree approximation as being due to the finite width
of the wave functions. 
To check this idea, we consider a ``quasi-Hartree'' approximation
(QHA) in which, instead of solving the problem   of Eq.~(\ref{eq:ham}),
we replace the wavefunctions by the eigenfunctions
(\ref{eq:Hermite-fu}) of the unconfined Landau problem and take the energy
eigenvalues from  Eq.~(\ref{eq:TFA-energies}). The latter would be
correct in the sense of a lowest order perturbation approximation with
respect to the effective confining potential $V(x)$, if $V(x)$ would
be a linear function of position over the extent of the unperturbed
wavefunction $\phi^0_{n,X}(x)$. The numerical effort with this QHA is
much less than that required for the full Hartree approximation, since
no numerical calculation of energy eigenvalues and -functions is
necessary. Density profiles calculated within this QHA are also
shown in Fig.~\ref{fig:incomp_compare} as dotted lines. It is seen
that the results are very similar to those of the full Hartree
calculation, in particular also the results for the smearing-out of
the incompressible strips. Apparently the smearing effect of the QHA
is even stronger than that of the full Hartree approximation. This is
understandable, since the Hartree wavefunctions are asymmetrically
squeezed in space regions of a rapid variation of $V(x)$, and
therefore have a smaller spatial extent than the unperturbed Landau
wavefunctions. 

In Fig.~\ref{comp-width-inc}
we compare the widths of incompressible strips with  $\nu(x)=2$
for several approximations. The line labeled CSG is the
analytical half-plane result of Chklovskii {\em et al.}
\cite{Chklovskii92:4026}, $\bar{a}_{\nu_{ic}}=16\sqrt{a_0 d_0/\pi}
\sqrt{\nu_{ic} /[\nu_0^{\,2}-\nu_{ic}^{\,2}]}$, with $d_0 \sim
150\,$nm a depletion length, $\nu_{ic} \,(=2)$ the filling factor of
the incompressible strip, and $\nu_0=2E_F^0/\hbar \omega_c$ the effective
filling factor. \cite{Lier94:7757} This result agrees well with the
corresponding self-consistent result of Ref.~\onlinecite{Lier94:7757}
and for sufficiently low $B$ values (note the inverted $B$ scale in
Fig.~\ref{comp-width-inc}) also with our present TFPA result
for samples of finite width. For small $B$ values, the width decreases
proportional to $B$, and remains finite throughout the figure. 

The result calculated from our QHA is also included in
Fig.~\ref{comp-width-inc}. As for the TFPA, we have determined the
width of the incompressible strips by a simple extrapolation, using
three points next to a plateau to determine a plateau edge.
 For wide plateaus (large $B$ values), the QHA width is
only slightly smaller than that calculated within the TFA. However,
with decreasing $B$ the QHA width decreases faster and goes to zero at
a relatively large magnetic field ($\hbar \omega_c/E_F^0 \approx
0.7$), far before the incompressible strips with $\nu(x)=4$ develop in
the center of the sample.

These results require a modification of Fig.~\ref{fig:sum-TFA}.
Within the Hartree-type approximation, the width of the incompressible
strip with filling factor 2 shrinks more rapidly with decreasing $B$
and vanishes at $\hbar \omega_c/E_F^0 \approx 0.7$. Between this $B$
value and the value $\hbar \omega_c/E_F^0 \approx 0.5$ no
incompressible strips exist in the sample. At still lower $B$ values
there is a $B$-interval in which only incompressible strips with
local filling factor 4 can exist. This modification required by the
QHA is indicated in Fig.~\ref{fig:result} below.

In view of the following it may be interesting to note that the
essential effects of the Hartree-type approximations can be simulated
in a very simple way. If we first perform a calculation within the TFPA 
and then take a spatial average, e.g.,
$\bar{\nu}(x)=(2\lambda)^{-1}\int_{-\lambda}^{\lambda} dx' \nu(x+x')$,
of the filling factor profile $\nu(x)$, we will smear-out
incompressible strips of a width less than $2\lambda$, while
incompressible strips with a width larger than  $2\lambda$ will survive.
With $\lambda$ of the order of the magnetic length, we will  obtain
filling factor profiles $\bar{\nu}(x)$  very similar to those obtained in 
the Hartree approximation. 



\section{Incompressible strips and distribution of dissipative current
  \label{sec:Conduc}} 

\subsection{The local conductivity model \label{seq:loc-cond}}

We now describe the effect of an applied current on our Hall bar
system, following again the approach of Ref.~\onlinecite{Guven03:115327}.
In the presence of a dissipative current $I$, the electrochemical
potential $\mu^{\star}({\bf r})$ will become position dependent, and
its gradient ${\bf E}={\bf \nabla}\mu^{\star}/e$ will drive the
current density {\bf j}({\bf r}). We assume the linear local relation
(Ohm's law)  
\be \label{eq:Ohm}
{\bf j}(x)=\hat{\sigma}(x)\, {\bf E}(x), \quad
\hat{\sigma}(x)=\hat{\sigma}\big(n_{\rm el}(x)\big),
\ee
with a position-dependent conductivity tensor $\hat{\sigma}(x)$, which
has the same form as for a homogeneous sample, however with the
homogeneous density replaced by the local electron density $n_{\rm el}(x)$.
Due to the translation invariance in the $y$ direction, which is indicated
in Eq.~(\ref{eq:Ohm}), and the equation of continuity, the components
$j_x$ and $E_y$ of current density and electric field, respectively,
must be constant,\cite{Guven03:115327} 
\be j_{x}(x)\equiv 0, \quad 
E_{y}(x) \equiv E_{y}^{0}\, . \label{eq:const-comp}
\ee
For the other components one finds
\be j_{y}(x)=\frac{1}{\rho_{l}(x)}E_{y}^{0}\,, \quad \label{eq:jy_x}
E_{x}(x)=\frac{\rho_{H}(x)}{\rho_{l}(x)}E_{y}^{0}\, , \ee
in terms of the longitudinal component $\rho_{l}=\rho_{xx}=\rho_{yy}$
and the Hall component 
$\rho_{H}=\rho_{xy}=-\rho_{yx}$  of the  resistivity tensor
$\hat{\rho}=\hat{\sigma}^{-1}$.
For a given applied current $I=\int_{-d}^d dx \,j_y(x)$ one obtains for
the constant electric field component along the Hall bar
\be \label{eq:e0y}
E_y^0= I \left[ \int_{-d}^d dx \, \frac{1}{\rho_l(x)} \right]^{-1} \,,
\ee
and for the Hall voltage across the sample
\be \label{eq:V_H}
V_{H}=\int_{-d}^{d}dx \,
E_{x}(x)=E_{y}^{0}\int_{-d}^{d}dx\, \frac{\rho_{H}(x)}{\rho_{l}(x)}\,.
\ee
With the usual normalization of the resistances to a square-shaped
conductor, this yields for the Hall and the longitudinal resistance
\be  \label{eq:resistances}
 R_{H}=\frac{V_{H}}{I} \,, \quad R_{l}=\frac{2dE_{y}^{0}}{I}\,.
\ee

Here we see the essence of the local model. Any reasonable model for
the conductivity of a high-mobility 2D ES at zero temperature will
give simple results for the conductivity components at even-integer
filling factors (where no elastic scattering is possible):
\be \label{eq:condateven-nu}
&&\sigma_l(\nu \mbox{=}2k)=\rho_l(\nu  \mbox{=}2k)=0\,, \\
&&\sigma_H(\nu \mbox{=}2k)=\frac{1}{\rho_H(\nu \mbox{=}2k)} =
\frac{e^2}{h}\, 2k \,. 
\ee
Thus, if an incompressible strip of finite width exists in the sample,
the integral in Eq.~(\ref{eq:e0y}) diverges and $E_y^0$ and,
therefore, the longitudinal resistance $R_l$ is zero. 
At low temperatures, $k_BT\ll\hbar \omega_c$, $\rho_l(\nu \mbox{=}2k)$ and,
therefore, $R_l$ will be exponentially small and
relevant contributions to the integral come only from the
incompressible regions. 

The  integral in Eq.~(\ref{eq:V_H}) has the same type of singularity. 
If only
incompressible strips with the same value $\nu(x)=2k$ of the local
filling factor exist, this singular integral is just the $2k$-fold of
the integral in Eq.~(\ref{eq:e0y}), so that we get the quantized result
$R_H=h/(2ke^2)$. At zero temperature one can evaluate the singular integrals 
by first introducing a cutoff, e.g. by replacing $\rho_l(x)$ with
$\rho_l^{\epsilon}(x)= \max[\epsilon, \rho_l(x)]$, then calculating the
integrals, and finally removing the cutoff ($\epsilon \rightarrow
0$). This yields exact quantization of the Hall resistance, and the
corresponding calculation at very low temperatures yields exponentially
small corrections. If
incompressible strips of finite widths with different values of
$\nu(x)$ exist, e.g. due to a manipulation of the background
potential, other values for the Hall resistance may be possible. But,
as we have learned from the Hartree-type approximations in the
previous section,  such a situation will not occur in our  simple
translation-invariant Hall bar geometry. From these arguments we
expect in the resistance-versus-B curves plateau regions of finite
widths, where the resistances have the well known quantized values.

These considerations are quite general and do not depend on details of
the conductivity model. On the other hand, if we want to calculate the
resistances between the plateau regions, we need to specify a
conductivity model, and the results will depend on details of this
model.
We will present such detailed results in Sec.~\ref{sec:Discuss} below.

\subsection{Limitations of the local model \label{sec:Lim-loc}}
In Sec.~\ref{sec:Hartree} we have shown that, within a Hartree-type
approximation, incompressible strips of a width smaller than the
extent of typical wavefunctions cannot exist. As a consequence,
incompressible strips with a given filling factor $2k$ do exist only in a
finite interval of magnetic fields. For lower $B$ values, the $\nu(x)$
profile crosses the value $2k=\nu(\tilde{x}_{2k})  $ with finite slope
at some point $x= \tilde{x}_{2k}$. At zero temperature, the integrals
in Eqs.~(\ref{eq:e0y}) and (\ref{eq:V_H}) become singular since
$\rho_l(  \tilde{x}_{2k})=0$. Whether the singularity is integrable or
not depends on the filling-factor-dependence of the longitudinal
conductivity, $\sigma_l(\nu)$. For the SCBA model, to be considered
below, it is integrable, for the Gaussian model considered by GG it is
not. But should we worry about such sophisticated questions depending
on details of the conductivity model? We think we should not, for the
following reasons.

All quantities that are related by Eq.~(\ref{eq:Ohm}), the current
density, the conductivity and the gradient of the electrochemical
potential, represent local values of physical variables, which are
defined by macroscopic statistical arguments. In principle, they have
to be calculated as average values over sufficiently small subsystems,  
which nevertheless should contain many electrons. We can not expect
that the local relation (\ref{eq:Ohm}) still holds on a length scale
of the order of the mean distance between the constituents of our 2D
ES, or, equivalently, of the order of the Fermi wavelength $\lambda_F$. 
On such a length scale one should consider a non-local version of
Ohm's law instead. This would, however, make things much more
complicated, and we will not enter such problems.

In order to simulate qualitatively the expected effects of a non-local
treatment, we start as before with a local model for the conductivity
tensor, take the spatial average over a length scale of the order of
$\lambda_F$, e.g.\ with $\lambda=\lambda_F/2$ as
\be \label{eq:mean-conduct}
\hat{\bar{\sigma}}(x)=\frac{1}{2\lambda}\int_{-\lambda}^{\lambda}
d\xi \, \hat{\sigma}(x+\xi)\,,
\ee
and use still the local version  (\ref{eq:Ohm}) of Ohm's law, but now
with the averaged conductivity tensor (\ref{eq:mean-conduct}). As a
consequence, the resistivity components occurring in
Sec.~\ref{seq:loc-cond} have to be calculated by tensor inversion of
$\hat{\bar{\sigma}}(x)$. 

This simple simulation of non-local effects has several appealing aspects.
First, if $\sigma_l(x)$ vanishes at an isolated position
$x=\tilde{x}_{2k}$, the averaged $\bar{\sigma}_l(x)>0$ will be
positive in the neighborhood of $\tilde{x}_{2k}$, and the integrals in
 Eqs.~(\ref{eq:e0y}) and (\ref{eq:V_H}) will not be
 singular. Intervals of vanishing $\bar{\sigma}_l(x)$ will exist only,
 if we start before averaging with sufficiently wide incompressible
 strips (wider than $2 \lambda$). Finally, for high-mobility systems,
 the Hall conductivity is given to a very good approximation by the
 free-electron value $\sigma_H(x)=(e^2/h)\nu(x)$. Thus, the averaged
 Hall conductivity $\bar{\sigma}_H(x)$ will be given by the averaged
 filling factor profile $\bar{\nu}(x)$. As mentioned at the end of
 Sec.~\ref{sec:Hartree}, this averaged profile will agree
 qualitatively with the Hartree profile, if we start with the TFPA
 profile $\nu(x)$ and average over $\lambda \sim l$ ($l$ the magnetic length).
Since for the large magnetic fields of our interest $l \lesssim
\lambda_F$, there is actually no need to perform the time consuming
Hartree calculation, if we finally want to calculate the averaged
conductivity tensor (\ref{eq:mean-conduct}).

To summarize, our approximation scheme that simulates both, the effect
of finite width of the wavefunctions in the thermal equilibrium
calculation, and non-local effects on the transport, is as follows. We
calculate the density profile $\nu(x)$ within the self-consistent
TFPA, and with this the conductivity tensor $\hat{\sigma}\big(\nu(x)\big)$.
Then we perform the averaging of Eq.~(\ref{eq:mean-conduct}) and
follow the calculations described in Sec.~\ref{seq:loc-cond}.
In contrast to Ref.~\onlinecite{Guven03:115327} we restrict our
calculations here to the linear response regime and do not investigate
the feedback of the applied current on the electron density and the
electrostatic potential, that is mediated in principle by the
position-dependent electrochemical potential $\mu^{\star}({\bf r})$ in
the presence of a dissipative current.

\subsection{Self-consistent Born approximation \label{sec:SCBA}}
In principle we could use the conductivity models of
Ref.~\onlinecite{Guven03:115327} in order to calculated explicit
examples. We prefer, however, to take the transport coefficients from
the self-consistent Born approximation (SCBA),
\cite{Ando74:959,Ando75:279,Ando82:437}  which allows for 
consistent models of longitudinal and Hall conductivities, and for the
consideration of anisotropic scattering by randomly distributed
finite-range impurity potentials. We assume that the relevant
scatterers are charged donors distributed randomly in a plane parallel
to that of the 2D ES, with an areal density $n_I$,
 and we approximate the impurity
potentials by  Gaussian potentials
\be \label{a:gauss-pot}
v({\bf r})=\frac{V_0}{\pi R^2}\, \exp\left(- \frac{r^2}{R^2}\right)\,,
\ee
with a range $R$ of the order of the
the spacing between 2D ES and doping layer.

An important aspect of the SCBA is that, similar to the ``lowest order
cumulant approximation'' used in Ref.~\onlinecite{Guven03:115327}, it
allows us to treat the transport coefficients and the collision
broadening of the LLs in a consistent manner.  The relevant
SCBA results for the transport coefficients and the collision broadening
 of a homogeneous 2D ES are summarized in the appendix. 
Consistency with the transport coefficients requires that we replace
in the self-consistent TFPA calculations the $\delta$-functions of the
Landau DOS (\ref{landau-dos}) by the semi-elliptic spectral functions
(\ref{a:spectral-fu}). 
In addition to the range $R$, the impurity strength $V_0$ and
concentration $n_I$ determine these quantities via the relaxation time
$\tau_0$ defined by the energy $\hbar/\tau_0= n_I V_0^2m/\hbar^2$. In
strong magnetic fields, this energy combines with the cyclotron energy
to
\be  \label{eq:gamsq}
\Gamma^2 =4 n_I V_0^2/(2\pi l^2)=(2/\pi)\, \hbar\omega_c \,
\hbar/\tau_0,
\ee
where $\Gamma$ is the width of the LLs in the limit of
zero-range scattering potentials ($R\rightarrow 0$). We find it
convenient to characterize the impurity strength by the dimensionless
ratio $\gamma =\Gamma /  \hbar\omega_c$ at the strong magnetic field
$B=10\,$T and define (for GaAs), therefore, the strength parameter
\be \label{eq:gamma_I}
\gamma_I =\big[(2n_I V_0^2m/\pi
  \hbar^2)/(17.3\,\mbox{meV})\big]^{1/2}\,.
\ee

\begin{figure}[ht]
{\centering
\includegraphics[width=0.78\linewidth,angle=-90]{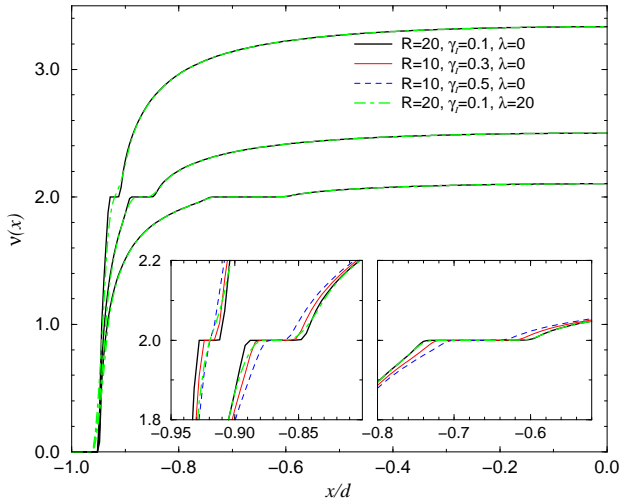}
\caption{Filling factor $\nu(x)\approx h\, \sigma_H(x)/e^2$
 versus position in the left half of a
  symmetric high-mobility ($R=20\,$nm, $\gamma_{I}=0.1$)
Hall bar of width $d=1.5\,\mu$m, for three values of the
  magnetic field, $\hbar \omega_c/E_F^0=0.6,\, 0.8,$ and 0.95, and
  without ($\lambda=0$, solid black lines) and with ($\lambda=20\,$nm,
  long-dashed green lines) averaging according to Eq.~(\ref{eq:mean-conduct}).
The insets show the plateau regions (incompressible strips) enlarged
 and include in addition two results for larger collision broadening
 but no averaging ($\lambda=0$). Other specifications in the text.
 \label{fig:plateaus}}}
\end{figure}
Figure.~\ref{fig:plateaus} shows the effect of collision broadening
on the density profile at strong magnetic fields. The sample
parameters are $d=1.5\,\mu$m, $n_0=4\cdot 10^{11}$cm$^{-2}$ and
$b/d=0.952$, resulting in 
$\bar{n}_{\rm el}=3.37\cdot 10^{11}$cm$^{-2}$ and  $E_F=12.02\,$meV,
$E_F^0= 13.51\,$meV.  Data are shown
for $t=k_BT/E_F^0=0.01$, three values of the magnetic field ($\hbar
\omega_c /E_F^0=0.6, \, 0.8$, and 0.95, corresponding to central
filling factors $\nu(0)=3.33, \, 2.5$, and 2.1, respectively), and for
three sets of the impurity parameters $R$ and $\gamma_{I}$. It is seen
from Fig.~\ref{fig:plateaus} and Table~\ref{table1}
that, for sufficiently small collision broadening (small $\gamma_{I}$
and, eventually, large $R$), incompressible strips still may exist, that
their width decreases, however, with increasing broadening of the
LLs. Table~\ref{table1} shows, for several sets of impurity
parameters, the relative widths $\gamma_n$ of the lowest LLs
and the zero field mobilities. Data for the second set ($R=10\,$nm,
$\gamma_I=0.1$) are not included in   Fig.~\ref{fig:plateaus}, since
they could not be distinguished from the traces for the first,
high-mobility set. From the insets of   Fig.~\ref{fig:plateaus} it is
evident that incompressible strips can only survive, if the gap
between the broadened LLs remains broad enough. A large
collision broadening (low-mobility set No. 4) has a similar effect as
a spatial averaging (long-dashed lines in   Fig.~\ref{fig:plateaus})
and may smear out the incompressible strips.

\begin{table}[h]
\centering
\begin{tabular}{c||c|c||l|l|l||c}
\hline
No. & $R$ [nm] & $\gamma_{I}$ & $\gamma_0$ & $\gamma_1$ & $\gamma_2$ 
& $\mu_{B=0}$  \\
\hline
1 & 20 & 0.1 & 0.07 & 0.06 & 0.05 & 747.5 \\
2 & 10 & 0.1 & 0.11 & 0.08 & 0.07 & 75.1 \\
3 & 10 & 0.3 & 0.34 & 0.24 & 0.21 & 8.34 \\
4 & 10 & 0.5 & 0.56 & 0.40 & 0.35 & 3.00 \\
\hline
\end{tabular}
\caption{Relative width $\gamma_n =\Gamma_n/\hbar \omega_c$ of the
  Landau levels $n=0,\,1,\,2$ at $\hbar \omega_c/E_F^0=0.6$, and
  mobility $\mu_{B=0}$ at $B=0$, $T=0$ in units of m$^2$/Vs, for four
  sets of model parameters $R$, 
  $\gamma_{I}$. \label{table1}}
\end{table}

%
\section{Results and discussion  \label{sec:Discuss}}
\subsection{Effect of spatial averaging \label{sec:spat_av}}

\begin{figure}[h]
{\centering
\includegraphics[width=0.85\linewidth,angle=-90]{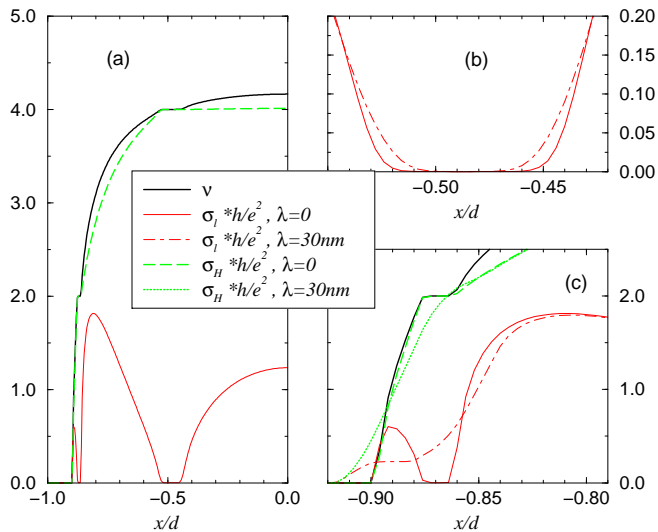}
\caption{(a) Filling factor and conductivity profiles for the left
  half of a
  symmetric sample with $d=1.5\,\mu$m, $n_0=4\cdot 10^{11}$cm$^{-2}$, $b=0.9d$,
calculated within the SCBA, for $R=0.1\,$nm and $\gamma_I=0.1$,
  $\hbar\omega_c/E_F^0=0.48$, and  $k_BT/E_F^0=0.01$.  (b) and (c)
  repeat the data of 
  (a) (solid lines) near the incompressible strips with filling
  factors $\nu(x)=4$ and $\nu(x)=2$, respectively. The dash-dotted
  lines demonstrate the effect of spatial averaging, according to
  Eq.~(\ref{eq:mean-conduct}), with $\lambda=30\,$nm.
 \label{fig:nu-sh-sl}}}
\end{figure}

The effect of spatial averaging, introduced to
simulate non-local effects on the scale of the Fermi wavelength, is
illustrated in Fig.~\ref{fig:nu-sh-sl}. It shows, for a magnetic
field value corresponding to a central filling factor $\nu(0)=4.18$,
the filling factor profile calculated with the SCBA broadened DOS,
together with the conductivity profiles,
Fig.~\ref{fig:nu-sh-sl}(a). Here we have assumed a short-range
potential (leading to the rather low mobility $\mu_{B=0}=6.4\,$m$^2$/Vs), in
order to obtain a noticeable deviation of the $\sigma_H(x)$ trace from that for
the filling factor $\nu(x)$.  Clear incompressible strips with the
quantized values for $\sigma_H(x)$ and vanishing $\sigma_l(x)$ are
visible where $\nu(x)$ assumes the integer values 4 and 2. The effect
of spatial averaging on the conductivities is demonstrated in (b) and (c). 
The wide ($\sim 90\,$nm) plateau defined by $\nu(x)=4$ shrinks  due to the
averaging (to $\sim 50\,$nm) but clearly survives, as is shown in
Fig.~\ref{fig:nu-sh-sl}(b) for 
$\sigma_l(x)$, and hold similarly for the plateau of $\sigma_H(x)$.  
On the other hand, the plateau behavior near the narrow   ($\sim 25\,$nm)
strip defined by  $\nu(x)=2$ is completely smeared out, and
$\sigma_l(x)$ does no longer vanish in this region, Fig.~\ref{fig:nu-sh-sl}(c).
This has, of course, drastic consequences for the current
distribution, which is dominated by the strips with vanishing
$\sigma_l(x)$, i.e., vanishing $\rho_l(x)$, see
Eq.~(\ref{eq:jy_x}). Without averaging a finite fraction of the total
current would flow through the incompressible strips  with $\nu(x)=2$
(on both sides of the sample). With the averaged conductivity tensor,
the total current must flow through the incompressible strips with
$\nu(x)=4$ (assuming that there $\sigma_l(x)=0$ holds exactly).

The mechanism illustrated in Fig.~\ref{fig:nu-sh-sl} is, of course,
also effective at other values of the magnetic field. At  larger $B$ with
$\nu(0) \gtrsim 2$, broad incompressible strips with $\nu(x)=2$ will
exist near the 
center of the Hall bar, and the spatially averaged conductivities will
show clear plateau behavior. With decreasing $B$, the incompressible
strips move from the center towards the sample edges and shrink.  If
the strip width  becomes of the order of $2 
\lambda$ or smaller, the averaging according to
Eq.~(\ref{eq:mean-conduct}) will destroy the plateau behavior of the
conductivities and $\bar{\sigma}_l(x)$ will no longer vanish near the
strips. Then the current density still may have a (finite) maximum
near the strips with  $\nu(x)=2$, but a finite amount of current will
spread over the bulk of the sample, and the global resistances will no
longer have the quantized values $R_H=h/2e^2$ and $R_l=0$.

\begin{figure}[ht]
{\centering
\includegraphics[width=0.85\linewidth,angle=-90]{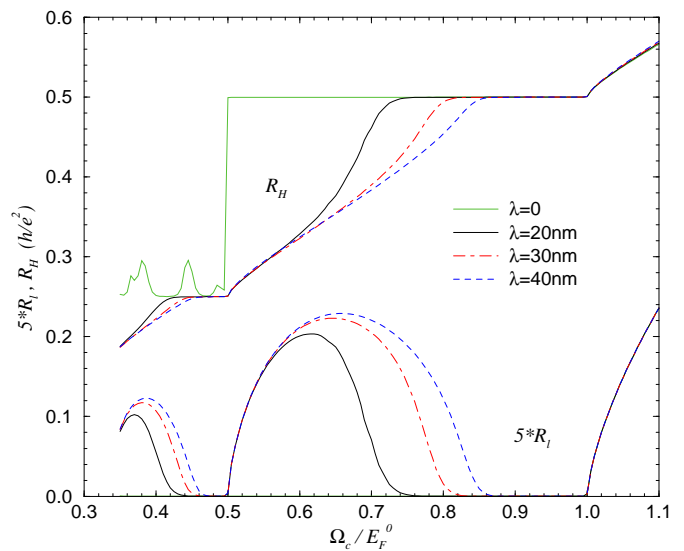}
\caption{Calculated Hall  and longitudinal resistances
  versus scaled magnetic field $\hbar\omega_c/E_F^0$, for
  different values of the averaging length $\lambda$. The sample
  parameters are  $d=1.5\,\mu$m, $n_0=4\cdot 10^{11}$cm$^{-2}$, $b=0.952d$,
 $R=10\,$nm and $\gamma_I=0.1$, and  $k_BT/E_F^0=0.01$. 
 \label{fig:RH5Rl_lambdas}}}
\end{figure}

Figure~\ref{fig:RH5Rl_lambdas} shows typical results for the
 dependences of the global resistances on magnetic field and averaging
 length, as  calculated from Eq.~(\ref{eq:resistances}).  
At high magnetic fields, $\hbar\omega_c>E_F^0$, we have everywhere in
 the sample $\nu(x)<2$, and no incompressible strips exist. The
 filling factor and consequently the conductivities and the current density
vary slowly across  the sample. Thus, the spatial
 averaging has little effect and, within the accuracy of the figure,
 the results with and without averaging agree. For slightly lower
 magnetic fields,  $\hbar\omega_c \lesssim E_F^0$, broad
 incompressible strips exist near the center of the sample and, for all
 the considered averaging lengths $\lambda$, strips of finite width with
 $\sigma_l(x)=0$ and $\sigma_H(x)=2e^2/h$ survive. As a consequence,
 the resistances are quantized, independently on the width of these
 strips.

For still lower $B$ values the situation becomes more
 complicated. Within the TFPA, incompressible strips exist for all
 these $B$ values. Without spatial averaging, $\sigma_l(x)$ vanishes on
 these strips and, as a consequence, $R_l=0$, and, for $\hbar\omega_c
 / E_F^0>0.5$, $R_H=h/2e^2$, as shown by the traces for
 $\lambda=0$. For   $\hbar\omega_c / E_F^0 <0.5$ there are two types
 of incompressible strips, with $\sigma_H(x)=2e^2/h$ or $4e^2/h$, and,
 without averaging,  both contribute according to their widths to
 $R_H$, while still $R_l=0$. The fluctuations in the $R_H$ curve for
 $\lambda=0$ result from our TFPA calculation on a finite mesh
 ($N=500$), which yields discontinuous changes of the widths of the
 incompressible strips with changing $B$.
This unsatisfactory picture, obtained
for $\lambda=0$, results from the  model calculation of
 Ref.~\onlinecite{Guven03:115327}. 

\begin{figure}[ht]
{\centering
\includegraphics[width=0.9\linewidth]{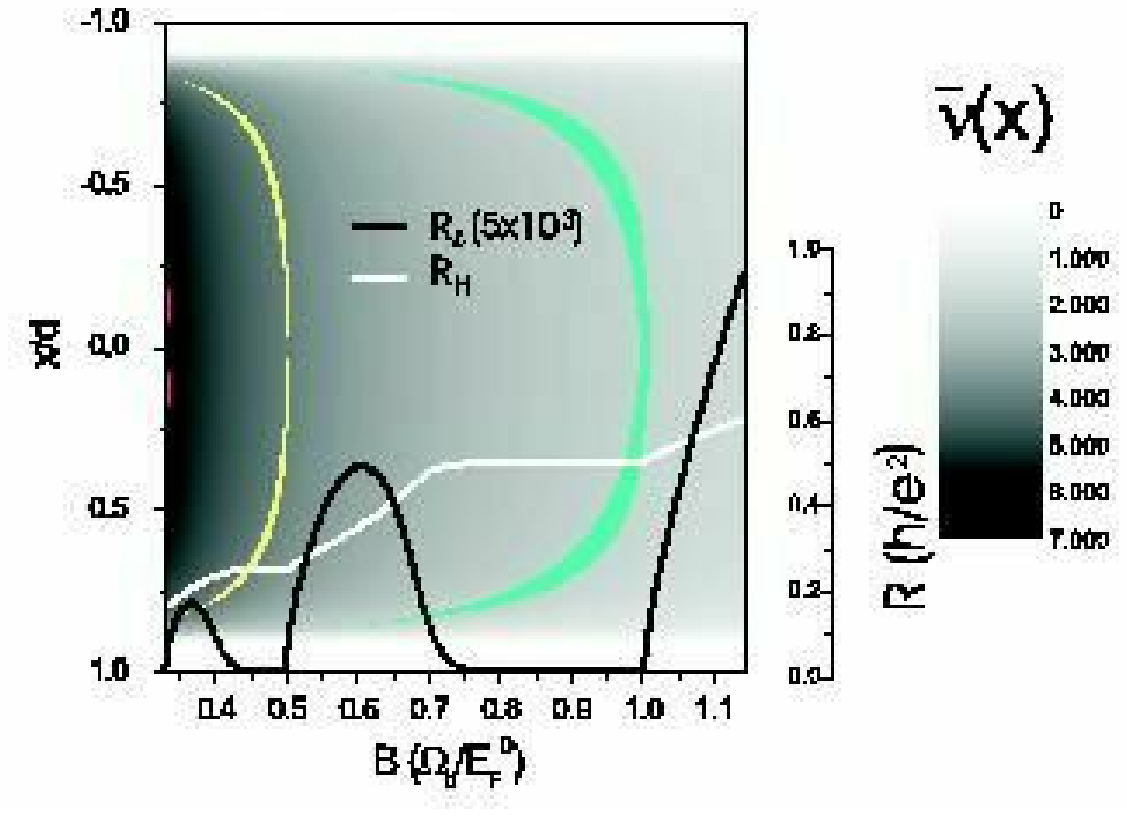}
\caption{Calculated Hall resistance (light solid line) and (scaled)
  longitudinal resistance (black solid line) versus magnetic field,
  measured in units of $\hbar \omega_c/E_F^0$. The underlying gray
  scale plot shows the averaged filling factor profile, as in
  Fig.~\ref{fig:sum-TFA}. The crescent-like areas indicate the regions
  of incompressible strips with local filling factors 2 (right) and 4 (left).
 \label{fig:result}}}
\end{figure}
The introduction of the spatial averaging improves this situation
dramatically and leads to qualitatively correct results. With
decreasing $B$ the incompressible strips with $\nu(x)=2$ become
narrower. As the width becomes smaller than $2\lambda$,
$\bar{\sigma}_l(x)$ no longer vanishes, the integrals in
Eqs.~(\ref{eq:e0y}) and (\ref{eq:V_H}) and thus $R_l$
 become finite. This happens at higher $B$ values if $\lambda$ is
 larger, and the resistances near the low-magnetic-field edge of the
 quantum Hall plateau depend strongly on  $\lambda$. While
 $\bar{\sigma}_l(x)$ may have a sharp minimum near the strips with
 $\nu(x)=2$ if the width of these strips is only slightly smaller that
 $2\lambda$, this minimum, and also the corresponding maximum of the
 current density, will smoothen out as the width of the strips
 becomes much smaller than $2\lambda$. Then the total resistances will
 become nearly independent of $\lambda$, as is seen in
 Fig.~\ref{fig:RH5Rl_lambdas} for $0.5<\hbar\omega_c/E_F^0\lesssim 0.6$.
For $\hbar\omega_c/E_F^0\lesssim 0.5$, $\bar{\sigma}_l(x)$ vanishes
only within the incompressible strips with $\nu(x)=4$, but not in
strips with $\nu(x)=2$. As a consequence, we obtain again the exactly
quantized results for $R_H$ and $R_l$.

To visualize the intimate connection between the existence of
incompressible strips of finite width [now with constant $\bar{\sigma}_H(x)
  \approx e^2 \bar{\nu}(x)/h$], we show  in Fig.~\ref{fig:result} a
gray scale plot of the spatially averaged filling factor profile for a
relevant interval of magnetic fields, together with the resulting resistances.

%
\subsection{Effect of temperature and collision broadening}
The spatial averaging procedure is essential to obtain quantum Hall (QH)
 plateaus of finite width for the $R_l(B)$ and the $R_H(B)$ curves and
 to obtain the correct quantized values for $R_H(B)$ on the plateaus
 corresponding to filling factors larger than two. The width of the
 calculated QH plateaus does, however, not only depend on the
 averaging length  $\lambda$, but also on the temperature and on the
 broadening of the LLs due to the impurity scattering, since
 both affect the width of the incompressible strips. 
The effect of collision broadening on the width of incompressible
strips has been indicated in Fig.~\ref{fig:plateaus}. The temperature
effect has been investigated in Ref.~\onlinecite{Lier94:7757},
where it was shown that, in the absence of collision broadening,
i.e., on the basis of the bare Landau DOS (\ref{landau-dos}), the
incompressible strips have a finite width at zero temperature. At
finite, increasing temperatures, the width
shrinks (while the value of the filling factor remains exactly constant
within the remaining strip) until at a sufficiently high temperature
($k_BT\lesssim \hbar \omega_c/25 $ in  Ref.~\onlinecite{Lier94:7757})
the width collapses to zero. A similar result is expected in the
presence of collision broadening. In particular we expect that, within
the self-consistent TFPA based on the SCBA DOS, the
existence of an energy gap between two adjacent broadened LLs
will always lead to an incompressible strip, provided the temperature
is low enough. The necessary temperature will be the lower, the
narrower the gap is. But we will not discuss these questions in further 
details.

\begin{figure}[ht]
{\centering \vspace*{-1cm}
\includegraphics[width=0.8\linewidth,angle=-90]{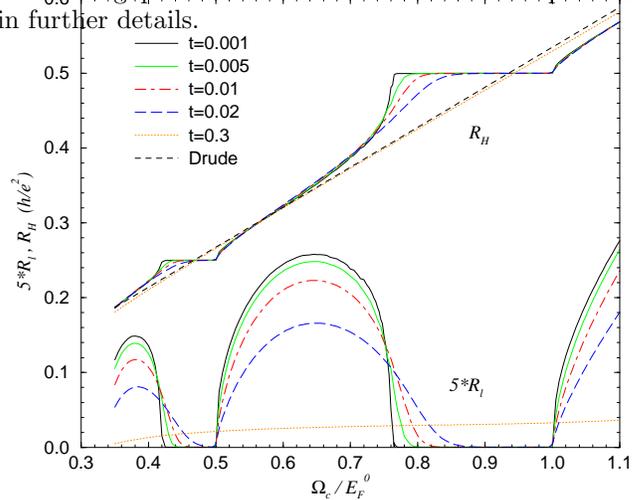}
\caption{ \label{fig:RlRH_s30_ts}
Hall and longitudinal resistances versus magnetic field, calculated
for different temperatures, $t=k_BT/E_F^0$. Sample parameters:
$d=1.5\,\mu$m, $n_0=4\cdot 10^{11}$cm$^{-2}$, $b=0.952d$, 
 $R=10\,$nm and $\gamma_I=0.1$, $\lambda=30\,$nm.
 }}
\end{figure}
The temperature effect on the calculated resistance curves is shown in
Fig.~\ref{fig:RlRH_s30_ts}. As expected, the width of the QH plateaus
increases monotonically with decreasing temperature, but apparently it
has a finite limit for $T\rightarrow 0$. We have also included the
high-temperature result for $k_BT/E_F^0=0.3$ (since $E_F^0=13.5\,$meV,
this means $T\approx 47\,$K).  At the low-$B$ side of the figure $k_BT
\sim \hbar \omega_c$, and the derivative of the Fermi function 
overlaps about two LLs. Since we consider only the lowest Landau
levels ($n=0,\, 1,\, 2$), our calculation is not correct in this limit.
Nevertheless it is interesting to compare this result with the Drude
result, which should be valid if the Shubnikov--de Haas oscillations
are smeared out at higher the temperature. 

In the Drude approximation we have $\rho_H(x)=\omega_c \tau_{tr} \,
\rho_l(x)$, with $\rho_l(x)= 1/ \sigma_0(x)$, where $ \sigma_0(x)=e^2
\tau_{tr} n_{\rm el}(x)/m$ is independent of the magnetic
field. Inserting this into Eq.~(\ref{eq:e0y}), we obtain 
\be \label{eq:Rl_Drude} 
\frac{I}{E_y^0} =\frac{e^2  \tau_{tr}}{m}\, 2d\,
\bar{n}_{\rm el} =\frac{e^2}{h}\, 
\frac{2 E_F}{\hbar / \tau_{tr}}\,.
\ee
In Eq.~(\ref{eq:V_H}) we take the integrand to be $\omega_c
\tau_{tr}$, but only for $|x|<b$,  where $n_{\rm el}(x)$ is not
exponentially small, and obtain  $V_H/E_y^0=2b\,
\omega_c\tau_{tr}$. With Eq.~(\ref{eq:resistances}) we obtain the
Drude result
\be
R_H^D=\frac{h}{e^2}\, \frac{b}{2d}\, \frac{E_F^0}{E_F}\,\frac{\hbar
  \omega_c}{E_F^0} \,, \quad R_l^D=\frac{h}{e^2}\, \frac{\hbar /
  \tau_{tr}}{2 E_F} \,.
\ee
The energies $E_F$ and $E_F^0$ are calculated numerically from the
density profile at $B=0$, $T=0$, as is described above, and  $
\tau_{tr}$ is calculated as described in the appendix, with
$k_F=\sqrt{2\pi  \bar{n}_{\rm el}}$. The results are plotted as dashed
 straight lines in Fig.~\ref{fig:RlRH_s30_ts}.

\begin{figure}[ht]
{\centering 
\includegraphics[width=0.8\linewidth,angle=-90]{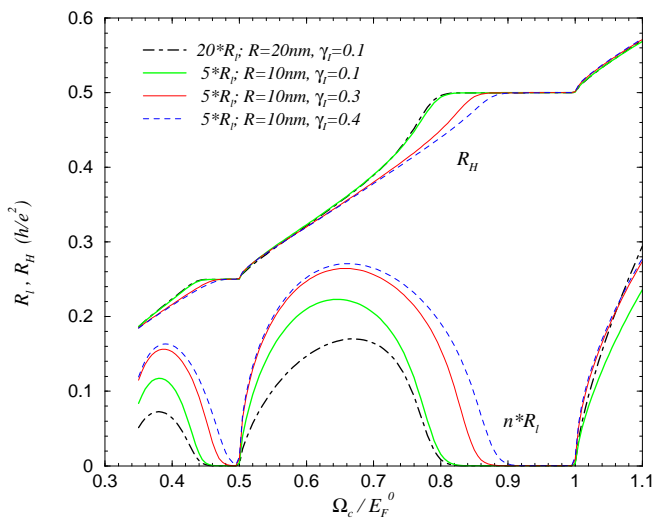}
\caption{ \label{fig:plat_mobil}
Hall and longitudinal resistances versus magnetic field, calculated
for different values of the collision broadening. Sample parameters:
$d=1.5\,\mu$m, $n_0=4\cdot 10^{11}$cm$^{-2}$, $b=0.952d$, 
 $\lambda=30\,$nm, $k_BT/E_F^0=0.01$ (i.e., $T=1.57\,$K). 
 }}
\end{figure}

Finally, Fig.~\ref{fig:plat_mobil} shows the effect of the Landau
level broadening on the QH plateaus at a fixed, relatively low temperature.
The corresponding widths of the lowest LLs, in units of the
cyclotron energy $\hbar \omega_c$, are given in Table~\ref{tab:2} 
for the lowest and the
largest $B$ value shown in the figure. For the largest damping, the
LLs start to overlap for $\hbar\omega_c /E_F^0 \lesssim
0.4$. so that the $\nu=4$ QH plateau is not well developed. We note
that the SCBA results summarized in the appendix are only valid at
higher $B$ values, where the LLs do not overlap. 

The two high-mobility situation considered in
Fig.~\ref{fig:plat_mobil} differ only by the range $R$ of the Gaussian
impurity potentials. The larger range leads to slightly smaller level
broadening, but to much lower longitudinal resistance (i.e., to much
higher mobility at $B=0$).

\begin{table}[h]  \centering
\begin{tabular}{c|c||c|c|c||c|c|c||c}
\hline
$R$ & $\gamma_I$ & $\gamma_0^l $ &  $\gamma_1 ^l $ &  $\gamma_2^l $ &
 $ \gamma_0^h $ &  $\gamma_1^h $ &  $\gamma_2^h $ & $\mu_{B=0}$\\
\hline
20 & 0.1 & 0.117 & 0.085 & 0.073 & 0.043 & 0.037 & 0.033 & 747.5\\
10 & 0.1 & 0.161 & 0.123 & 0.105 & 0.071 & 0.051 & 0.044 & 75.1\\
10 & 0.3 & 0.482 & 0.369 & 0.316 & 0.213 & 0.152 & 0.131 & 8.34\\
10 & 0.4 & 0.643 & 0.492 & 0.421 & 0.284 & 0.203 & 0.175 & 4.69\\
\hline
\end{tabular}
\caption{\label{tab:2} Relative widths of the lowest Landau levels,
  $\gamma_n^l =\Gamma_n/\hbar\omega_c$ for $\hbar\omega_c/E_F^0=0.35$,
and $\gamma_n^h =\Gamma_n/\hbar\omega_c$ for
  $\hbar\omega_c/E_F^0=1.10$, for the impurity parameters used in
  Fig.~\ref{fig:plat_mobil}. First column: range in nm, last column:
  zero field mobility in m$^2$/Vs.
}
\end{table}

\subsection{Hall potential profile}
The motivation of Ref.~\onlinecite{Guven03:115327} and our present
work came from the experimental investigation \cite{Ahlswede01:562} of
the electrostatic potential distribution across a Hall bar under QH
conditions, caused by an applied current. Ahlswede and coworkers
\cite{Ahlswede01:562,Ahlswede02:165} observed three types of potential
distribution, depending on the filling factor regime. Type I was a
more or less linear variation across the sample and is observed if the
filling factor in the center is smaller and relatively close to (but
not too close to) an
integer $n$, i.e., $n\gtrsim \nu(0) \gtrsim n-1/2$. If the center
filling is slightly larger than an integer, $n < \nu(0) \lesssim
n+1/2$, type III is observed,  characterized by  a  constant
potential in the central region and a rapid variation across (narrow)
strips, which move with decreasing $B$ towards the sample edges and
have been interpreted as incompressible strips.\cite{Ahlswede01:562}
Finally, type II shows a rapid, non-linear variation of the potential
in the center region and is observed if the center filling factor is
very close to an integer.  

In Ref.~\onlinecite{Guven03:115327} it was shown that, in an local
equilibrium picture, the changes of the electrostatic potential, caused
by an applied current, follows closely the current-induced variation of
the electro-chemical potential $\mu^{\star}$, so that the resulting
density changes are small. In the present work we do not consider the
feed-back of the spatial variation of $\mu^{\star}$ on 
electrostatic potential and density profiles (linear response). But we
expect from the results of Ref.~\onlinecite{Guven03:115327}, that, in
the linear response regime, $\mu^{\star}$  should show the same
position dependence as the electrostatic potential would do, if the feed-back
were calculated.

\begin{figure}[ht]
{\centering 
\includegraphics[width=\linewidth]{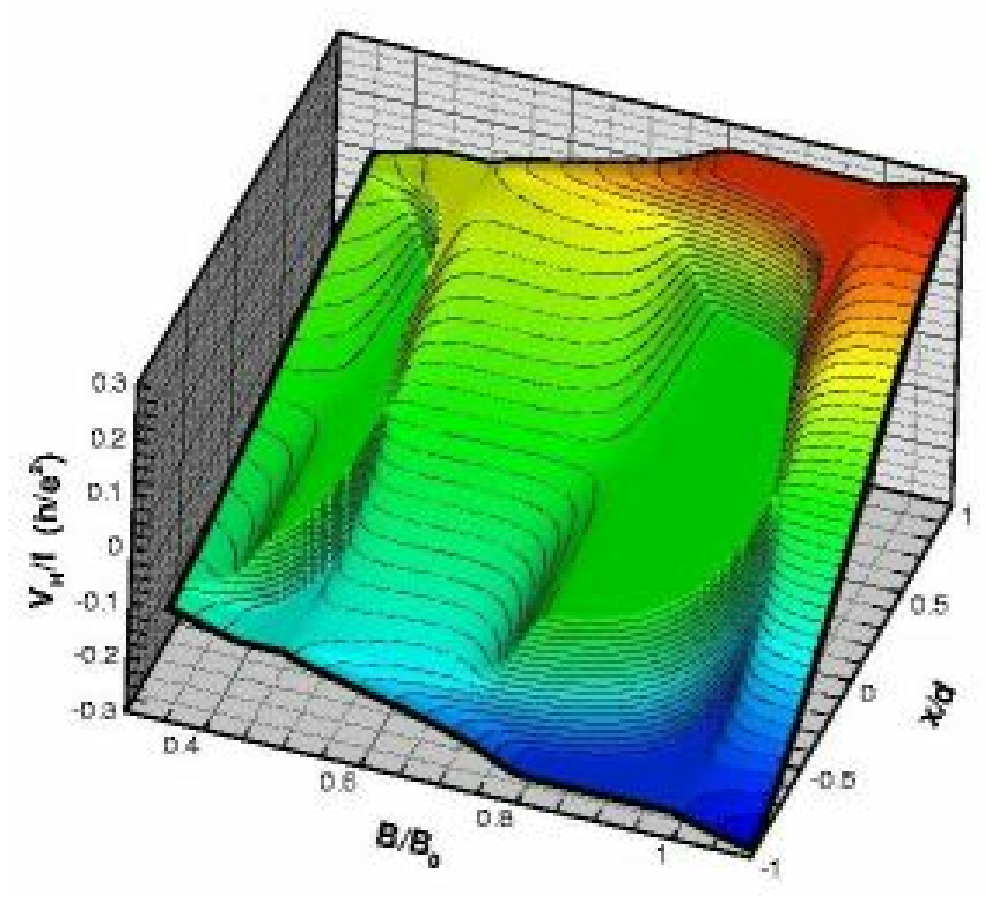}
\caption{ \label{fig:elch_3d}
Hall potential profile $V_H(x)=\int_0^x dx'E_x(x')$ across the sample, for
varying $B/B_0\equiv 
\hbar\omega_c/E_F^0$ and constant applied current $I$.  Normalization:
$[V_H(d)-V_H(-d)]/I=R_H$; sample parameters:
$d=1.5\,\mu$m, $n_0=4\cdot 10^{11}$cm$^{-2}$, $b=0.952d$, 
 $\lambda=30\,$nm, $k_BT/E_F^0=0.01$. 
 }}
\end{figure}

To calculate the Hall profile across the sample, we integrate
$E_x(x)$, Eq.~(\ref{eq:jy_x}), from the center $x=0$ to the actual $x$
value. Typical results as functions of position $x$ and magnetic field
$B$ are shown in Fig.~\ref{fig:elch_3d}. The normalization is chosen
so that $V_H(B,x=d)/I=-V_H(B,x=-d)/I=R_H/2$.
One sees clearly that the plateaus of the quantized Hall effect ($0.8
\lesssim \hbar\omega_c /E_F^0 \lesssim 1$ and $0.45 \lesssim
\hbar\omega_c /E_F^0 \lesssim 0.5$) coincide with potential variation
of type III, caused 
by current density confinement to the incompressible strips. Moving
from  a plateau region to smaller $B$ values, the incompressible
strips shrink and finally vanish, and the current density spreads more
and more out into the bulk. This leads to the type I behavior ($0.52    
\lesssim \hbar\omega_c  /E_F^0  \lesssim 0.7$ and $\hbar\omega_c
/E_F^0 \lesssim 0.4$). Immediately above the integer values of the
center filling factor (in our approximation assuming spin degeneracy
near $\nu(0)=2$ and 4), we find the rapid variation of the type II.
This is  in very nice agreement with the
experiment. Without our spatial averaging of the
conductivity tensor, we would have  missed the type I regions for
$\hbar\omega_c /E_F^0  <1$, as has been observed in
Ref.~\onlinecite{Guven03:115327}. 

\subsection{Summary} 
The virtue of our approach is, that it allows us to calculate
resistance traces with exactly quantized quantum Hall plateaus of
finite width, and with reasonable values of the resistances between
these plateaus. While these intermediate resistance values depend on
the details of our conductivity model, the quantized plateau values do
not. The reason for this high accuracy and model-independence of the
plateau values is the fact, that the latter are determined by
the integrals  in Eqs.~(\ref{eq:e0y}) and (\ref{eq:V_H}) becoming
singular across incompressible strips.

To obtain realistic widths of the QH plateaus, we had to consider a
mechanism that prohibits singular current flow along very narrow
incompressible strips. We have argued that small-scale non-local
transport effects act into this direction, and that consideration of
the finite extent of wavefunctions will prohibit arbitrarily narrow
ISs at low magnetic fields, in contrast to the  prediction of the
Thomas-Fermi approximation. We were able to simulate such non-local
effects by a simple spatial averaging procedure,  with reasonable results
for Hall and longitudinal 
resistance as functions of the magnetic field. Also the resulting
potential profile, and therefore the current distribution across the
sample, is in nice agreement with recent investigations.\cite{Ahlswede01:562} 
We consider this as a strong support for the relevance of our approach,
notably because earlier approaches, which neglected dissipation, cannot
explain the experiments, as has been discussed in
Ref.~\onlinecite{Guven03:115327}.

Note that, for QH plateaus corresponding to filling factors $\nu \geq 4$,
our results are qualitatively different form the conventional edge
channel picture. The latter explains, for instance, the quantized conductance
value $G=4\,e^2/h$ as the sum of the contributions of two
spin-degenerate, quasi-one-dimensional  current channels near each of
the opposite sample edges, thus tracing
back the quantized Hall effect to the phenomenon of 1D conductance
quantization (in a situation where no backscattering
occurs). \cite{Buettiker88:9375}  That is,
the edge states, created by the LLs with quantum numbers
$n=0$ and $n=1$, contribute both to the current in the plateau regime
of the QHE. Our results, on the contrary, indicate that the total current
flows along the incompressible strip with local filling factor
$\nu(x)=4$ (where both LLs $n=0$ and $n=1$ are occupied), whereas near
local filling factor $\nu(x)=2$ no incompressible strip and no
contribution to the current exists.

Comparing our resistance curves with experiments, we notice that the high-field
edge of a calculated plateau occurs at a magnetic field, at which an
incompressible strip (with an even integer value of the effective
filling factor $\nu_0=2E_F^0/\hbar \omega_c$) first occurs in the
center of the sample. In experiments these $\nu_0$
values usually are found somewhere near the centers of the plateaus.
We have good arguments that this discrepancy is due to our neglect of
long-range potential fluctuations due to the randomly distributed
ionized donors. We have simulated the ``short-range'' part of the
Coulomb potentials of the remote donors  by Gaussian
potentials, but we have neglected their overlapping long-range parts,
which lead to long-range potential fluctuations.
We have evaluated the short-range disorder within the SCBA to
calculate conductivities and LL broadening. We have seen
that with increasing disorder scattering the level broadening
increases and, as a result, the widths of the QH plateaus shrinks. On
the other hand, one knows from technical applications of the QH
effect, that rather impure samples have usually especially
wide and stable QH plateaus. This points to the role of long-range
potential fluctuations, which become more important with increasing
impurity concentration. 

As a rough simulation of such long-range fluctuations, we have added 
oscillatory terms to the confinement potential and than repeated our
calculations. We indeed find that such modulations can widen and
stabilize the QH plateaus, and eventually even shift them to higher
magnetic fields, depending on the amplitude, the range, and possible
other details of the perturbation. This becomes understandable, if one
considers the effect of such fluctuations on the existence and the
position of
incompressible strips. However, we do not want to discuss such
considerations further, since we believe that our
quasi-one-dimensional model is not appropriate for a reliable
discussion of statistical fluctuations of a 2D donor
distribution. Effects of long-range disorder in an unconfined 2D ES on
the longitudinal 
resistance between QH plateaus have already been discussed a decade
ago.\cite{Chklovskii93:18060} This discussion seems however not
applicable to the rather narrow samples of our present interest,
since, first, the confinement affects the self-consistently calculated
potential and thus the density distribution, \cite{Siddiki03:125315}
and, second, the early assumptions about current-carrying and
isolating regions are not compatible with our results and the
experimental findings.\cite{Ahlswede01:562} 

\acknowledgments
We gratefully acknowledge useful discussions with E.~Ahlswede and J.~Weis.
This work was supported by the Deutsche
Forschungsgemeinschaft, SP ``Quanten-Hall-Systeme'' GE306/4-2.

\appendix*
\section{SCBA conductivities}
The low-temperature, high-field magnetotransport, determined
by elastic scattering of the 2D electrons by randomly distributed
impurities with scattering potentials of arbitrary range, has been
studied by Ando and coworkers. \cite{Ando74:959,Ando75:279,Ando82:437} 
The results for the case of non-overlapping LLs can be summarized as
\begin{figure}[h]
{\centering
\includegraphics[width=\linewidth]{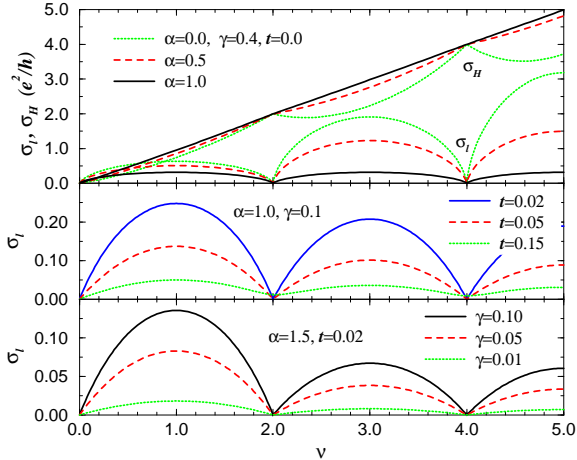}
\caption{SCBA results for
longitudinal  ($\sigma_l$) and  Hall ($\sigma_H$) conductivity, in
units of $e^2/h$, versus
filling factor,  at fixed magnetic field
 for different values of impurity range ($\alpha=R/l$) and strength
 ($\gamma=\Gamma/\hbar \omega_c$), and of temperature ($t=k_BT/\hbar
 \omega_c$). For the two lower panels the correction $-\Delta \sigma_H$ to
 $\sigma_H^0=e^2\nu /h$ is negligible, and $\sigma_H$ is not shown.
 \label{fig:condpar}}}
\end{figure}
%
\be
\nu&=&g_s\sum_{n=0}^{\infty}\int dE\,A_n(E)\, f(E-\mu)\,,
\label{a:nu}\\
\sigma_l&=& g_s\sum_{n=0}^{\infty}\int dE\left[-\frac{\partial
    f}{\partial E}\right]  \, \sigma_{xx}^{(n)}(E)\,, \label{a:sigma_l}\\
\sigma_H&=& \frac{e^2}{h}\,\nu -\Delta \sigma_H \,,\label{a:sigma_H}\\
\Delta \sigma_H&=& g_s\sum_{n=0}^{\infty}\int dE\left[-\frac{\partial
    f}{\partial E}\right]  \,\Delta \sigma_{yx}^{(n)}(E)\,, \label{a:Dsigma_H}
\ee
with the spectral functions  of widths $\Gamma_n$, 
\be \label{a:spectral-fu}
A_n(E)=\frac{2}{\pi \Gamma_n} \,
\sqrt{1-\left(\frac{E-E_n}{\Gamma_n}\right)^2}\,,
\ee
centered around the Landau
energies (\ref{eq:TFA-energies}), and
\be \label{a:slvonE}
 \sigma_{xx}^{(n)}(E)&=&\frac{e^2}{h}\,\frac{\pi}{2}\,
 \Big[\Gamma_n^{xx}\, A_n(E)\Big]^2\,, \\
\Delta \sigma_{yx}^{(n)}(E)&=&\frac{e^2}{h}\,\frac{\pi^2}{4}
\frac{\Gamma_n^{yx}}{\hbar \omega_c}\, \Big[\Gamma_n^{yx}\, A_n(E)\Big]^3\,.
\ee
 Assuming a single type of impurities
 with the Gaussian potential (\ref{a:gauss-pot}), these parameters can
 be expressed in terms of the integrals
\be
(\Gamma_n^{(j)})^2 =\Gamma^2 \int_0^{\infty}dx\,g_n^{(j)}(x) \,
\exp(-[1+\alpha^2]x) \,,
\ee
where $\Gamma^2=4n_IV_0^2/(2\pi l^2)$ and $\alpha =R/l$, and the
weight functions 
\be
 g_n^{(0)}(x)&=&\Big[L_n^0(\alpha^2 x)\Big]^2\,,\quad
 g_n^{(d)}(x)= \frac{1-x}{2 \alpha^2}\, g_n^{(0)}(x)\,,\nonumber\\
g_n^{(\pm)}(x)&=&\frac{x}{\sqrt{2n+1\pm 1}}\, L_n^0(\alpha^2 x)
L^1_{n-(1\mp1)/2}(\alpha^2 x)\nonumber
\ee
are determined by the associated Laguerre polynomials $L_{n}^{m}(x)$.
With these notations one obtains
\be 
&&\Gamma_n^2 = (\Gamma_n^{(0)})^2\,, \quad 
(\Gamma_n^{xx})^2 = (\Gamma_n^{(d)})^2\,,\\[0.1cm]
&& (\Gamma_n^{yx})^4 =(\Gamma_n^{(+)})^4+(\Gamma_n^{(-)})^4\,.
\ee\\[-0.5cm]

In the limit of short-range scattering potentials, $\alpha \rightarrow 0$,
one has $\Gamma_n^2 /\Gamma^2=1 $, 
$~(\Gamma_n^{xx}/\Gamma )^2 =n+1/2 $
and $(\Gamma_n^{yx}/ \Gamma)^4 =n+1/2 $. With increasing $\alpha$
these parameters decrease and remain for $\alpha \gtrsim 1$ about an
order of magnitude smaller than their $\alpha=0$
values. \cite{Ando74:959,Ando75:279} 

Some typical SCBA results  are shown in Fig.~\ref{fig:condpar}. 
In general, $\sigma_l$ and the correction $\Delta \sigma_H$ to the
free electron Hall conductivity $\sigma_H^0=e^2\nu /h$ decrease with
increasing range of the scattering potentials.

 At zero temperature,  
$\Delta \sigma_H$ is proportional to $\Gamma/\hbar \omega_c$ with a
factor depending only on the range parameter $\alpha=R/l$. The
longitudinal conductivity $\sigma_l(\nu)$, on the other hand, depends
only on $\alpha$ and not on $\Gamma^2=(2/\pi) \hbar\omega_c\,
\hbar/\tau_0$, i.e. depends not on the impurity concentration $n_I$
and strength $V_0$ entering the $B=0$ relaxation rate $\hbar/\tau_0=
n_I V_0^2m/\hbar^2$. This is very different from the $B=0$
conductivity $\sigma_0 =e^2 n_{\rm el} \tau_{tr}/m$ obtained for the
impurity model (\ref{a:gauss-pot}), which depends on
 $n_{\rm el}=k_F^2/2\pi$ and via 
 $\tau_{tr}$ on both the potential strength and range. For $B=0$
collision broadening effects can be neglected, and one obtains for
elastic scattering at the Fermi egde
\be
\frac{\hbar}{\tau_{tr}}&=& \frac{n_I m}{\hbar^2}\int_{-\pi}^{\pi}
\frac{d\varphi}{2\pi} \big[v_q\big]^2_{q=k_F[2 (1-\cos
    \varphi)]^{1/2}}\, (1-\cos  \varphi) \nonumber \\
&=& \frac{\hbar}{\tau_0} \, \big[e^{-x} ( I_0(x)-I_1(x))\big]_{x=(Rk_F)^2}\,,
\ee
where the last equality holds for our impurity potential
(\ref{a:gauss-pot}), with Fourier transform $v_q=V_0 \exp(-R^2
q^2/4)$, and  $I_{\nu}(x)$ is a modified Bessel function. This  leads for $Rk_F
\gg 1$ to $\tau_0/\tau_{tr} \approx [\sqrt{8\pi}\, (Rk_F)^3]^{-1}$.
With increasing temperature, the peak values of $\sigma_l(\nu)$
decrease and the minima at even integer $\nu$ values are no longer
exponentially small for $k_BT/\hbar \omega_c \gtrsim 0.1$. This
behavior of the SCBA results, shown in the middle panel of
Fig.~\ref{fig:condpar}, is similar to that of the Gaussian model
treated in Ref.~\onlinecite{Guven03:115327}.  At finite temperature,
the longitudinal conductivities $\sigma_l(\nu)$ increase with
$\gamma$, i.e., with increasing scattering strength (bottom panel of
Fig.~\ref{fig:condpar}). This is as expected from the Drude picture for
$\omega_c \tau_{tr}\gg 1$: $\sigma_l \approx \sigma_0/(\omega_c
\tau_{tr})^2 \propto 1/\tau_{tr}$.


\end{document}